\documentclass[aip,jcp,reprint]{revtex4-1}
\usepackage{graphicx}

\begin{document}

\title{Fine discretization of pair interactions and an approximate analytical strategy for predicting equilibrium behavior of complex fluids}

\author{Kyle B. Hollingshead}
\affiliation{McKetta Department of Chemical Engineering, University of Texas at Austin, Austin, Texas 78712, USA}

\author{Avni Jain}
\affiliation{McKetta Department of Chemical Engineering, University of Texas at Austin, Austin, Texas 78712, USA}


\author{Thomas M. Truskett}
\email[]{truskett@che.utexas.edu}
\affiliation{McKetta Department of Chemical Engineering, University of Texas at Austin, Austin, Texas 78712, USA}

\date{\today}

\begin{abstract}
We study whether fine discretization (i.e., terracing) of continuous pair interactions, when used in combination with first-order mean-spherical approximation theory, can lead to a simple and general analytical strategy for predicting the equilibrium structure and thermodynamics of complex fluids. 
Specifically, we implement a version of this approach to predict how screened electrostatic repulsions, solute-mediated depletion attractions, or ramp-shaped repulsions modify the radial distribution function and the potential energy of reference hard-sphere fluids, and we compare the predictions to exact results from molecular simulations.
\end{abstract}


\maketitle

\section{Introduction}
Effective interactions between suspended colloids or nanoparticles can be experimentally tuned (e.g., by changing the properties of the solvent, through physical or chemical modification of the particles, or via external fields) so that macroscopic properties of the corresponding complex fluids 
can be engineered from the ``bottom up.''\cite{yethiraj:tunedcolloids,likos:tunedcolloids,Glotzer2007,Chen2011}
Statistical mechanics provides a formal quantitative framework that links the relevant microscopic and macroscopic properties, in principle allowing for computational inverse design of interparticle interactions to achieve desired material characteristics ({\em e.g.,} specific structural features or other targeted properties via structure-property relations).
\cite{torquato:invstatmech,PhysRevLett.100.106001,C1SM06932B, marcotte:invstatmech2, cohn:invstatmech, edlund:invstatmech,marcotte:061101,jain:invstatmech}

In practice, successful inverse design strategies rely upon on accurate means for solving a forward version of the problem at hand. Molecular simulations or sophisticated integral equation theories--otherwise well suited for the forward calculation of equilibrium behavior from microscopic interactions--currently require computational resources that are prohibitive for use in most optimization strategies.  Simple analytic liquid-state theories are a potentially attractive alternative, but they are unfortunately limited in terms of the types of pair potentials that they can 
accurately treat.\cite{thysimpliq} 
In this Communication, we explore whether fine discretization (i.e., terracing) of the pair interaction can allow one to use analytical theories in a new way to predict the behaviors of a broader range of model complex fluids, rendering these analytical methods more powerful as tools for materials design.

Our proposed strategy comprises three parts: (1) fine discretization of a short-range, continuous pair potential into a terraced representation, (2) application of an approximate, analytical liquid-state theory capable of accurately predicting the finely sawtoothed radial distribution function (RDF) of the terraced model, and (3) smoothing of the sawtoothed RDF to recover a continuous prediction for the pair correlation function of the fluid with the original potential. 

To test the performance of this discretization-and-smoothing based approach, we compare its predictions to exact (within numerical precision) results from molecular simulations.
Specifically, we study the accuracy of its predictions for how short-range screened electrostatic (Yukawa)\cite{davoudi:yukawa, cochran:yukawa, heinen:yukawa} repulsions, solute-induced depletion (Asakura-Oosawa)\cite{asakura:asakuraoosawa, roth:asakuraoosawa} attractions, or ramp-shaped\cite{yan:ramp, jagla:ramp, errington:ramp} repulsions modify the equilibrium structure and thermodynamics of a hard-sphere fluid.

\section{Methods}
\subsection{Discretization Strategy}
We consider isotropic, pairwise interparticle interactions $\varphi(r)$ that consist of a hard-core exclusion for separations less than a particle diameter ($r<\sigma$) plus a continuous, short-range contribution $\phi(r)$ that decays to zero by a cut-off $r_c$, 
\begin{equation}
\label{eq-potential}
\varphi\left(r\right) = \left\{ \begin{array}{ll}
\infty & ~~r < \sigma \\
\phi\left(r\right) & ~~\sigma \leq r < r_c \\ 
0 & ~~r \ge r_c
\end{array} \right..
\end{equation}

\begin{figure}
\includegraphics[width=8.5cm]{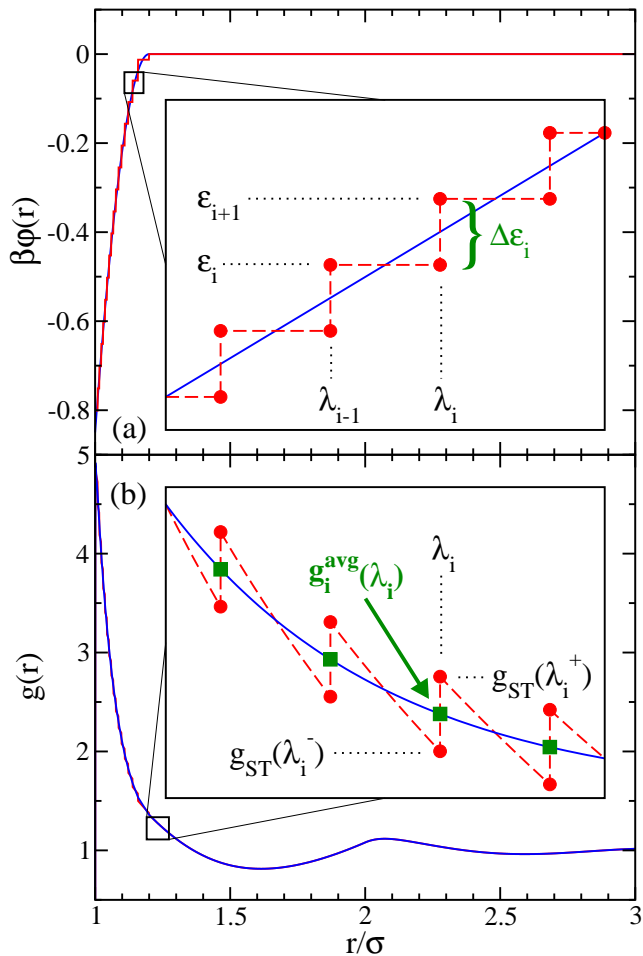}
\caption{\label{fig_example}Schematic of potential terracing and RDF smoothing.  (a) The continuous potential of interest $\beta \varphi(r)$ (blue, solid) and the corresponding terraced version (red, dashed) are shown, with parameters $\epsilon_i$ and $\lambda_i$ determined from Eq.~\ref{eq-terracing} as described in the text. (b) The sawtoothed RDF $g_{\text{ST}}(r)$ (red, dashed) associated with the terraced potential is computed using Eq.~\ref{eq-gST}. It is smoothed using Eqs.~\ref{eq_gofr} and \ref{eq-smooth-corr} to arrive at a continuous prediction (blue, solid) for the RDF of the fluid with the original potential.}
\end{figure}

We discretize the continuous potential into a terraced representation of $M$ steps, each with a constant energy 
\begin{equation}
\varepsilon_i = \left(\lambda_i-\lambda_{i-1}\right)^{-1}\int_{\lambda_{i-1}}^{\lambda_i}{\phi\left(r\right)dr} ,
\label{eq-terracing}
\end{equation}
where $\lambda_i$ is the outer boundary of step $i$ (see Figure \ref{fig_example}a).
The values of $\varepsilon_i$ and $\lambda_i$ are determined simultaneously, beginning from $r_c$ and working inward, such that the difference in energies between neighboring steps, $\Delta\varepsilon_i = \varepsilon_{i+1} - \varepsilon_i$, is smaller in magnitude than a specified threshold $\Delta\varepsilon^{max}$. 
In this work, we set $\beta \Delta\varepsilon^{max} = 0.05$, where $\beta=(k_{\text B} T)^{-1}$, $k_{\text B}$ is the Boltzmann constant and $T$ is temperature. 

Terraced potentials produce sawtoothed shaped RDFs, $g_{\text{ST}}\left(r\right)$, which we compute here via an extension of the simple exponential first-order mean spherical approximation (SEXP-FMSA) proposed by Hlushak, et al. for square-shoulder systems,\cite{hlushak:sexpfmsasquareshoulder} which itself is a variation of the first-order mean spherical approximation (FMSA) of Tang and Lu\cite{tang:fmsa2, tang:fmsa}.  
Specifically, we treat each step in the discretized potential as an independent perturbation to the RDF of the reference hard-sphere fluid 
at the same packing fraction~$\eta$:
\begin{equation}
g_{\text{ST}}\left(r\right)= g_{\text{HS}}\left(r\right)\prod_{i=1}^M{\exp\left[-\beta\Delta\varepsilon_ig_{\text{FMSA}}\left(r,\lambda_i,\eta\right)\right]}
\label{eq-gST}
\end{equation}
where $g_{\text{HS}}\left(r\right)$ is the RDF of the hard-sphere fluid, and $g_{\text{FMSA}}(r,\lambda_i,\eta)$ is the FMSA perturbation defined by Eq. 73 of Ref. \onlinecite{tang:fmsa2}. The quantity $g_{\text{FMSA}}(r,\lambda_i,\eta)$ depends on $\lambda_i$ and $\eta=\pi\rho\sigma^3/6$, where $\rho$ is the number density. This particular approximation is capable of treating terraced potentials with a cut-off $r_c \le 2 \sigma$, a constraint that might be relaxed in future implementations by choosing a different analytical theory for this step.

To arrive at a continuous prediction for the pair correlations of the fluid with the original potential, we smooth out the ``teeth'' in $g_{\text{ST}}(r)$ by adding a correction $\Delta g\left(r\right)$,
\begin{equation} \label{eq_gofr}
g\left(r\right)\equiv g_{\text{ST}}\left(r\right)+\Delta g\left(r\right), 
\end{equation} 
which is a piecewise sequence of linear functions:  
\begin{eqnarray}
\Delta g\left(r\right) \equiv && \left[g_i^{avg}-g_{\text{ST}}\left(\lambda_i^-\right)\right] \left(\frac{r-\lambda_{i-1}}{\lambda_i-\lambda_{i-1}}\right) \nonumber \\ &&+ \left[g_{i-1}^{avg}-g_{\text{ST}}\left(\lambda_{i-1}^+\right)\right] \left(1-\frac{r-\lambda_{i-1}}{\lambda_i-\lambda_{i-1}}\right)  \nonumber\\
&& {\text{for}} \quad \lambda_{i-1}<r<\lambda_i, 
\label{eq-smooth-corr}
\end{eqnarray} 
with $g_i^{avg} = \left[g_{\text{ST}}\left(\lambda_i^-\right)+g_{\text{ST}}\left(\lambda_i^+\right)\right] / 2$ .

\subsection{Model Pair Interactions}
As alluded to above, we consider three forms for $\phi(r)$ in Eq.~\ref{eq-potential}: screened electrostatic (Yukawa)\cite{davoudi:yukawa, cochran:yukawa, heinen:yukawa} repulsions, solute-induced depletion (Asakura-Oosawa)\cite{asakura:asakuraoosawa, roth:asakuraoosawa} attractions, or ramp-shaped\cite{yan:ramp, jagla:ramp, errington:ramp} repulsions.

\paragraph{Screened electrostatic repulsions.} The repulsive Yukawa potential can be expressed as
\begin{equation}
\phi_{\text{Y}}\left(x\right)=\gamma x^{-1} \exp \left[-\kappa\left(x-1\right)\right],
\end{equation}
where $x=r/\sigma$, $\gamma>0$ is the energy at contact (effective Yukawa charge), and $\kappa^{-1}$ is the screening length. To avoid discontinuities,
we adopt a form that is cut at $x_c = 2$ and shifted
\begin{equation}
\phi(x)=\phi_{\text{Y}}\left(x\right) - \phi_{\text{Y}}\left(x_c\right).
\end{equation}

\paragraph{Depletion attractions.} The Asakura-Oosawa (AO) model for non-interacting, solute-induced depletion attractions can be expressed
\begin{equation}
\beta \phi\left(x\right)=-\frac{\eta_p}{\left(1-x_c^{-1}\right)^3}\left[1-\frac{3}{2} \frac{x}{x_c} +\frac{1}{2} \left(\frac{x}{x_c}\right)^3\right]
\end{equation}
for $1 \leq x \leq x_c$ and $x_c = \left(1+q\right)$. Here, $q$ is the implicit solute to explicit particle diameter ratio, and $\eta_p$ is the packing fraction of implicit solute particles.

\paragraph{Ramp-shaped repulsions.} In its simplest form, the hard-core plus repulsive ramp potential is:
\begin{equation}
\phi\left(x\right) = U_1 \left[1-(x/x_c)\right]
\end{equation}
for $1 \leq x \leq x_c$. Here, $U_1$ is the characteristic energy scale of the ramp, and we choose $x_c=2$.

\begin{figure*}
\includegraphics[width=17cm]{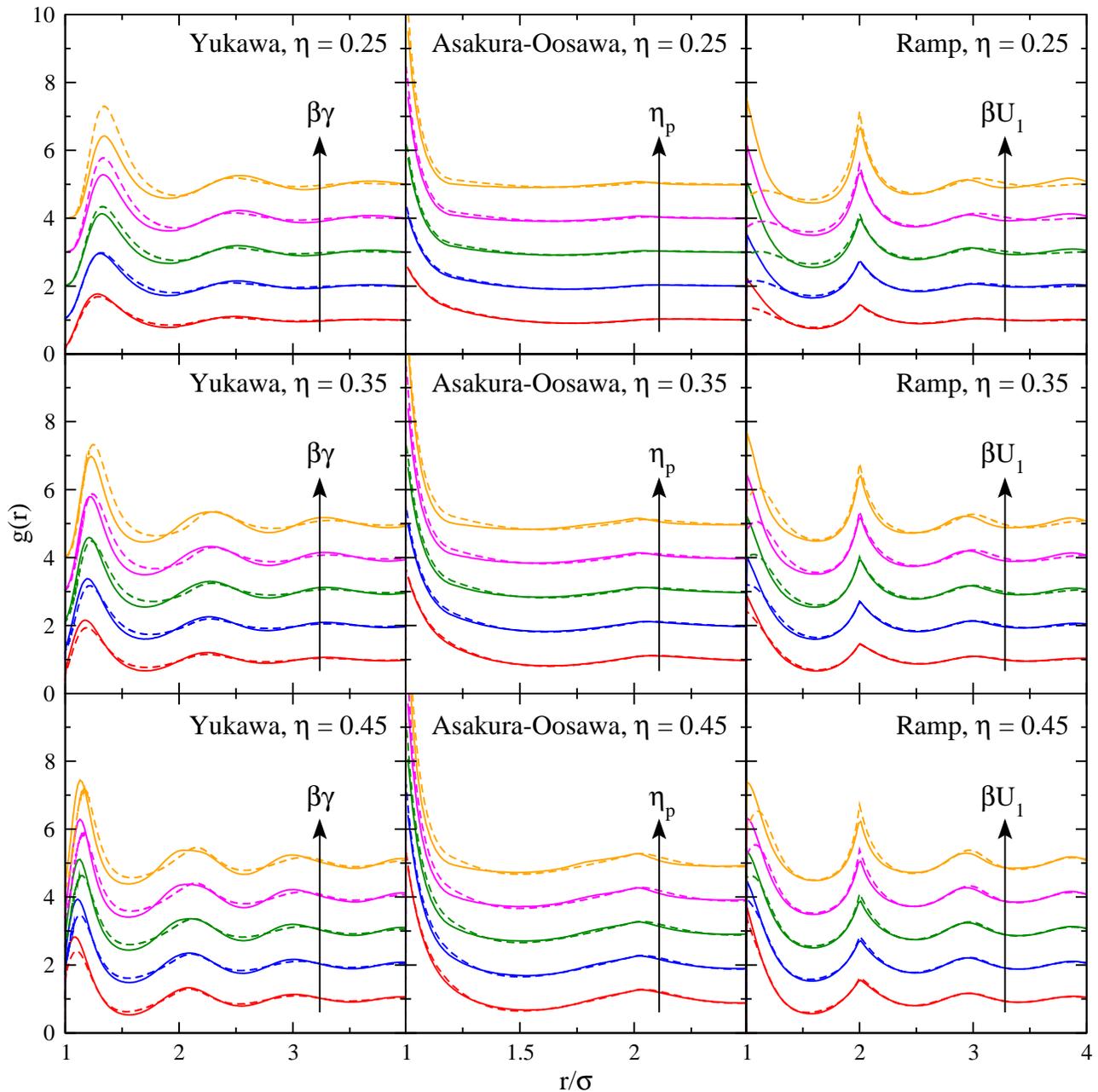}
\caption{\label{fig_gofrs}RDFs of fluids with particles interacting via hard-sphere plus Yukawa repulsions, Asakura-Oosawa (AO) depletion attractions, and ramp-shaped repulsions at packing fractions $\eta=0.25$, $0.35$, and $0.45$. Shown are predictions of Eq. (\ref{eq_gofr}) (dashed lines) and results of Monte Carlo simulations (solid lines). For the Yukawa repulsions, $\kappa=5$ and $\beta\gamma = 5$ (red), $7.5$ (blue), $10$ (green), $12.5$ (magenta), or $15$ (orange).  For the AO depletion attraction, $q=0.2$ and $\eta_p = 0.04$ (red), $0.08$ (blue), $0.12$ (green), $0.16$ (magenta), or $0.2$ (orange). For the repulsive ramp interaction, $r_c=2\sigma$ and $\beta U_1 = 10$ (red), $15$ (blue), $20$ (green), $25$ (magenta), or $30$ (orange). Curves for different interaction strengths are shifted vertically by integer values for clarity.}
\end{figure*}

\subsection{Molecular Simulations}
We test the discretization-and-smoothing strategy by comparing its smoothed RDF and potential energy predictions with exact results from canonical-ensemble Monte Carlo molecular simulations. We initialized the Monte Carlo simulations with either $N=2744$ particles (for systems with Yukawa repulsions) or $N=1000$ particles (for systems with AO depletion attractions or ramp-shaped repulsions) in disordered configurations within a cubic simulation box, using periodic boundary conditions. After an initial equilibration period at the temperature of interest, we collected properties over a period of $10^6$ Monte Carlo cycles. 



\section{Comparison of Analytical Predictions to Simulation Results}
To assess the performance of our proposed theoretical strategy, we investigate its ability to predict static structure (quantified by the RDF) and internal energy for the three model systems discussed above. 

The predicted RDFs of Eq.~\ref{eq_gofr} and those computed from simulations for a range of packing fractions ($\eta$ = 0.25 - 0.45) and potential interaction strengths are presented in Figure~\ref{fig_gofrs}. Broadly speaking, the predictions capture the simulated pair correlations of the three systems, despite the fact that each represent significant--and qualitatively different--departures from the structure of the hard-sphere fluid.  
The theoretical strategy predicts the most accurate structures for higher packing fractions and weaker interactions. The only qualitative failing appears in the strongly interacting repulsive-ramp fluid at low density, where the contact value of the RDF is significantly underpredicted.   

\begin{figure}
\includegraphics[width=8.5cm]{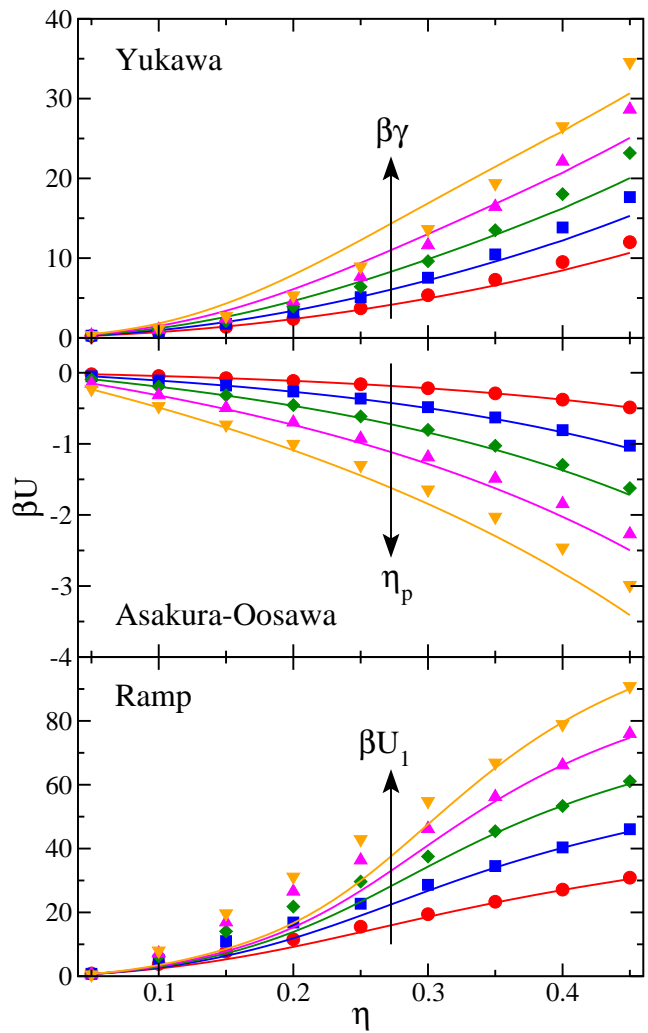}
\caption{\label{fig_us} Potential energy $\beta U$ versus packing fraction~$\eta$ for fluids with particles interacting via hard-sphere plus Yukawa repulsions, Asakura-Oosawa (AO) depletion attractions, and ramp-shaped repulsions. Solid curves are obtained from the predicted RDFs via $\beta U = (\beta/2) \int \varphi(r) g(r) d {\bf r}$, and symbols are results from Monte Carlo simulations. Interaction potential parameters for each system ($\beta\gamma$, $\phi_p$, and $\beta U_1$) are the same as in Fig. \ref{fig_gofrs}.}
\end{figure}

Potential energies from predicted by our strategy and those computed from simulations as functions of interaction strength and packing fraction are presented in Figure \ref{fig_us}.
In general, the trends track what might be expected from the RDFs shown in Figure~\ref{fig_gofrs}: very good agreement at high packing fractions, except at the highest potential interaction strengths, and also good agreement for low interaction strengths at all packing fractions (especially for the fluid with AO depletion attractions). The largest quantitative deviations of the theoretical predictions from the simulations occur for repulsive Yukawa fluid and the repulsive ramp model for strong energies of interaction in the packing fraction range ($\eta < 0.25$). 

\section{Conclusion}
In summary, the analytical, discretization-based approach we introduce here can predict the thermodynamic and structural consequences of some of the diverse types of short-range interactions that naturally emerge in dense, complex fluids (e.g., suspended colloids). Since good predictive strategies can provide guidance on how best to tune these systems to achieve desired changes in macroscopic properties, this approach represents a promising new strategy for using analytical liquid-state theories as tools for materials design.

\begin{acknowledgments}
The authors thank J. R. Solana for his generous assistance with the Tang-Lu FMSA solution and equations, and Jeffrey Errington for help with Monte Carlo simulation. 
T.M.T. acknowledges support of the Welch Foundation (F-1696) and the National Science Foundation (CBET-1065357). 
We also acknowledge the Texas Advanced Computing Center (TACC) at The University of Texas at Austin for providing HPC resources that have contributed to the research findings reported within this paper.
\end{acknowledgments}

\bibliography{articles}

\begin{thebibliography}{24}%
\makeatletter
\providecommand \@ifxundefined [1]{%
 \@ifx{#1\undefined}
}%
\providecommand \@ifnum [1]{%
 \ifnum #1\expandafter \@firstoftwo
 \else \expandafter \@secondoftwo
 \fi
}%
\providecommand \@ifx [1]{%
 \ifx #1\expandafter \@firstoftwo
 \else \expandafter \@secondoftwo
 \fi
}%
\providecommand \natexlab [1]{#1}%
\providecommand \enquote  [1]{``#1''}%
\providecommand \bibnamefont  [1]{#1}%
\providecommand \bibfnamefont [1]{#1}%
\providecommand \citenamefont [1]{#1}%
\providecommand \href@noop [0]{\@secondoftwo}%
\providecommand \href [0]{\begingroup \@sanitize@url \@href}%
\providecommand \@href[1]{\@@startlink{#1}\@@href}%
\providecommand \@@href[1]{\endgroup#1\@@endlink}%
\providecommand \@sanitize@url [0]{\catcode `\\12\catcode `\$12\catcode
  `\&12\catcode `\#12\catcode `\^12\catcode `\_12\catcode `\%12\relax}%
\providecommand \@@startlink[1]{}%
\providecommand \@@endlink[0]{}%
\providecommand \url  [0]{\begingroup\@sanitize@url \@url }%
\providecommand \@url [1]{\endgroup\@href {#1}{\urlprefix }}%
\providecommand \urlprefix  [0]{URL }%
\providecommand \Eprint [0]{\href }%
\providecommand \doibase [0]{http://dx.doi.org/}%
\providecommand \selectlanguage [0]{\@gobble}%
\providecommand \bibinfo  [0]{\@secondoftwo}%
\providecommand \bibfield  [0]{\@secondoftwo}%
\providecommand \translation [1]{[#1]}%
\providecommand \BibitemOpen [0]{}%
\providecommand \bibitemStop [0]{}%
\providecommand \bibitemNoStop [0]{.\EOS\space}%
\providecommand \EOS [0]{\spacefactor3000\relax}%
\providecommand \BibitemShut  [1]{\csname bibitem#1\endcsname}%
\let\auto@bib@innerbib\@empty
\bibitem [{\citenamefont {Yethiraj}\ and\ \citenamefont {van
  Blaaderen}(2003)}]{yethiraj:tunedcolloids}%
  \BibitemOpen
  \bibfield  {author} {\bibinfo {author} {\bibfnamefont {A.}~\bibnamefont
  {Yethiraj}}\ and\ \bibinfo {author} {\bibfnamefont {A.}~\bibnamefont {van
  Blaaderen}},\ }\href {\doibase {10.1038/nature01328}} {\bibfield  {journal}
  {\bibinfo  {journal} {{Nature}}\ }\textbf {\bibinfo {volume} {{421}}},\
  \bibinfo {pages} {{513}} (\bibinfo {year} {{2003}})}\BibitemShut {NoStop}%
\bibitem [{\citenamefont {Likos}(2001)}]{likos:tunedcolloids}%
  \BibitemOpen
  \bibfield  {author} {\bibinfo {author} {\bibfnamefont {C.~N.}\ \bibnamefont
  {Likos}},\ }\href {\doibase http://dx.doi.org/10.1016/S0370-1573(00)00141-1}
  {\bibfield  {journal} {\bibinfo  {journal} {Phys. Rep.}\ }\textbf {\bibinfo
  {volume} {348}},\ \bibinfo {pages} {267 } (\bibinfo {year}
  {2001})}\BibitemShut {NoStop}%
\bibitem [{\citenamefont {Glotzer}\ and\ \citenamefont
  {Solomon}(2007)}]{Glotzer2007}%
  \BibitemOpen
  \bibfield  {author} {\bibinfo {author} {\bibfnamefont {S.~C.}\ \bibnamefont
  {Glotzer}}\ and\ \bibinfo {author} {\bibfnamefont {M.~J.}\ \bibnamefont
  {Solomon}},\ }\href {\doibase 10.1038/nmat1949} {\bibfield  {journal}
  {\bibinfo  {journal} {Nature Materials}\ }\textbf {\bibinfo {volume} {6}},\
  \bibinfo {pages} {557} (\bibinfo {year} {2007})}\BibitemShut {NoStop}%
\bibitem [{\citenamefont {Chen}, \citenamefont {Bae},\ and\ \citenamefont
  {Granick}(2011)}]{Chen2011}%
  \BibitemOpen
  \bibfield  {author} {\bibinfo {author} {\bibfnamefont {Q.}~\bibnamefont
  {Chen}}, \bibinfo {author} {\bibfnamefont {S.~C.}\ \bibnamefont {Bae}}, \
  and\ \bibinfo {author} {\bibfnamefont {S.}~\bibnamefont {Granick}},\ }\href
  {\doibase 10.1038/nature09713} {\bibfield  {journal} {\bibinfo  {journal}
  {Nature}\ }\textbf {\bibinfo {volume} {469}},\ \bibinfo {pages} {381}
  (\bibinfo {year} {2011})}\BibitemShut {NoStop}%
\bibitem [{\citenamefont {Torquato}(2009)}]{torquato:invstatmech}%
  \BibitemOpen
  \bibfield  {author} {\bibinfo {author} {\bibfnamefont {S.}~\bibnamefont
  {Torquato}},\ }\href {\doibase 10.1039/B814211B} {\bibfield  {journal}
  {\bibinfo  {journal} {Soft Matter}\ }\textbf {\bibinfo {volume} {5}},\
  \bibinfo {pages} {1157} (\bibinfo {year} {2009})}\BibitemShut {NoStop}%
\bibitem [{\citenamefont {Goel}\ \emph {et~al.}(2008)\citenamefont {Goel},
  \citenamefont {Krekelberg}, \citenamefont {Errington},\ and\ \citenamefont
  {Truskett}}]{PhysRevLett.100.106001}%
  \BibitemOpen
  \bibfield  {author} {\bibinfo {author} {\bibfnamefont {G.}~\bibnamefont
  {Goel}}, \bibinfo {author} {\bibfnamefont {W.~P.}\ \bibnamefont
  {Krekelberg}}, \bibinfo {author} {\bibfnamefont {J.~R.}\ \bibnamefont
  {Errington}}, \ and\ \bibinfo {author} {\bibfnamefont {T.~M.}\ \bibnamefont
  {Truskett}},\ }\href {\doibase 10.1103/PhysRevLett.100.106001} {\bibfield
  {journal} {\bibinfo  {journal} {Phys. Rev. Lett.}\ }\textbf {\bibinfo
  {volume} {100}},\ \bibinfo {pages} {106001} (\bibinfo {year}
  {2008})}\BibitemShut {NoStop}%
\bibitem [{\citenamefont {Carmer}\ \emph {et~al.}(2012)\citenamefont {Carmer},
  \citenamefont {Goel}, \citenamefont {Pond}, \citenamefont {Errington},\ and\
  \citenamefont {Truskett}}]{C1SM06932B}%
  \BibitemOpen
  \bibfield  {author} {\bibinfo {author} {\bibfnamefont {J.}~\bibnamefont
  {Carmer}}, \bibinfo {author} {\bibfnamefont {G.}~\bibnamefont {Goel}},
  \bibinfo {author} {\bibfnamefont {M.~J.}\ \bibnamefont {Pond}}, \bibinfo
  {author} {\bibfnamefont {J.~R.}\ \bibnamefont {Errington}}, \ and\ \bibinfo
  {author} {\bibfnamefont {T.~M.}\ \bibnamefont {Truskett}},\ }\href {\doibase
  10.1039/C1SM06932B} {\bibfield  {journal} {\bibinfo  {journal} {Soft Matter}\
  }\textbf {\bibinfo {volume} {8}},\ \bibinfo {pages} {4083} (\bibinfo {year}
  {2012})}\BibitemShut {NoStop}%
\bibitem [{\citenamefont {Marcotte}, \citenamefont {Stillinger},\ and\
  \citenamefont {Torquato}(2011)}]{marcotte:invstatmech2}%
  \BibitemOpen
  \bibfield  {author} {\bibinfo {author} {\bibfnamefont {E.}~\bibnamefont
  {Marcotte}}, \bibinfo {author} {\bibfnamefont {F.~H.}\ \bibnamefont
  {Stillinger}}, \ and\ \bibinfo {author} {\bibfnamefont {S.}~\bibnamefont
  {Torquato}},\ }\href {\doibase 10.1063/1.3576141} {\bibfield  {journal}
  {\bibinfo  {journal} {J. Chem. Phys.}\ }\textbf {\bibinfo {volume} {134}},\
  \bibinfo {eid} {164105} (\bibinfo {year} {2011})}\BibitemShut {NoStop}%
\bibitem [{\citenamefont {Cohn}\ and\ \citenamefont
  {Kumar}(2009)}]{cohn:invstatmech}%
  \BibitemOpen
  \bibfield  {author} {\bibinfo {author} {\bibfnamefont {H.}~\bibnamefont
  {Cohn}}\ and\ \bibinfo {author} {\bibfnamefont {A.}~\bibnamefont {Kumar}},\
  }\href {\doibase 10.1073/pnas.0901636106} {\bibfield  {journal} {\bibinfo
  {journal} {Proc. Natl. Acad. Sci. U.S.A.}\ }\textbf {\bibinfo {volume}
  {106}},\ \bibinfo {pages} {9570} (\bibinfo {year} {2009})}\BibitemShut
  {NoStop}%
\bibitem [{\citenamefont {Edlund}, \citenamefont {Lindgren},\ and\
  \citenamefont {Jacobi}(2011)}]{edlund:invstatmech}%
  \BibitemOpen
  \bibfield  {author} {\bibinfo {author} {\bibfnamefont {E.}~\bibnamefont
  {Edlund}}, \bibinfo {author} {\bibfnamefont {O.}~\bibnamefont {Lindgren}}, \
  and\ \bibinfo {author} {\bibfnamefont {M.~N.}\ \bibnamefont {Jacobi}},\
  }\href {\doibase 10.1103/PhysRevLett.107.085503} {\bibfield  {journal}
  {\bibinfo  {journal} {Phys. Rev. Lett.}\ }\textbf {\bibinfo {volume} {107}},\
  \bibinfo {pages} {085503} (\bibinfo {year} {2011})}\BibitemShut {NoStop}%
\bibitem [{\citenamefont {Marcotte}, \citenamefont {Stillinger},\ and\
  \citenamefont {Torquato}(2013)}]{marcotte:061101}%
  \BibitemOpen
  \bibfield  {author} {\bibinfo {author} {\bibfnamefont {E.}~\bibnamefont
  {Marcotte}}, \bibinfo {author} {\bibfnamefont {F.~H.}\ \bibnamefont
  {Stillinger}}, \ and\ \bibinfo {author} {\bibfnamefont {S.}~\bibnamefont
  {Torquato}},\ }\href {\doibase 10.1063/1.4790634} {\bibfield  {journal}
  {\bibinfo  {journal} {The Journal of Chemical Physics}\ }\textbf {\bibinfo
  {volume} {138}},\ \bibinfo {eid} {061101} (\bibinfo {year}
  {2013})}\BibitemShut {NoStop}%
\bibitem [{\citenamefont {Jain}, \citenamefont {Errington},\ and\ \citenamefont
  {Truskett}(2013)}]{jain:invstatmech}%
  \BibitemOpen
  \bibfield  {author} {\bibinfo {author} {\bibfnamefont {A.}~\bibnamefont
  {Jain}}, \bibinfo {author} {\bibfnamefont {J.~R.}\ \bibnamefont {Errington}},
  \ and\ \bibinfo {author} {\bibfnamefont {T.~M.}\ \bibnamefont {Truskett}},\
  }\href {\doibase 10.1039/C3SM27785B} {\bibfield  {journal} {\bibinfo
  {journal} {Soft Matter}\ }\textbf {\bibinfo {volume} {9}},\ \bibinfo {pages}
  {3866} (\bibinfo {year} {2013})}\BibitemShut {NoStop}%
\bibitem [{\citenamefont {Hansen}\ and\ \citenamefont
  {McDonald}(2006)}]{thysimpliq}%
  \BibitemOpen
  \bibfield  {author} {\bibinfo {author} {\bibfnamefont {J.}~\bibnamefont
  {Hansen}}\ and\ \bibinfo {author} {\bibfnamefont {I.}~\bibnamefont
  {McDonald}},\ }\href {http://books.google.com/books?id=Uhm87WZBnxEC} {\emph
  {\bibinfo {title} {Theory of Simple Liquids}}}\ (\bibinfo  {publisher}
  {Elsevier Science},\ \bibinfo {year} {2006})\BibitemShut {NoStop}%
\bibitem [{\citenamefont {Davoudi}\ \emph {et~al.}(2000)\citenamefont
  {Davoudi}, \citenamefont {Kohandel}, \citenamefont {Mohammadi},\ and\
  \citenamefont {Tanatar}}]{davoudi:yukawa}%
  \BibitemOpen
  \bibfield  {author} {\bibinfo {author} {\bibfnamefont {B.}~\bibnamefont
  {Davoudi}}, \bibinfo {author} {\bibfnamefont {M.}~\bibnamefont {Kohandel}},
  \bibinfo {author} {\bibfnamefont {M.}~\bibnamefont {Mohammadi}}, \ and\
  \bibinfo {author} {\bibfnamefont {B.}~\bibnamefont {Tanatar}},\ }\href
  {\doibase 10.1103/PhysRevE.62.6977} {\bibfield  {journal} {\bibinfo
  {journal} {Phys. Rev. E}\ }\textbf {\bibinfo {volume} {62}},\ \bibinfo
  {pages} {6977} (\bibinfo {year} {2000})}\BibitemShut {NoStop}%
\bibitem [{\citenamefont {Cochran}\ and\ \citenamefont
  {Chiew}(2004)}]{cochran:yukawa}%
  \BibitemOpen
  \bibfield  {author} {\bibinfo {author} {\bibfnamefont {T.~W.}\ \bibnamefont
  {Cochran}}\ and\ \bibinfo {author} {\bibfnamefont {Y.~C.}\ \bibnamefont
  {Chiew}},\ }\href {\doibase 10.1063/1.1759616} {\bibfield  {journal}
  {\bibinfo  {journal} {J. Chem. Phys.}\ }\textbf {\bibinfo {volume} {121}},\
  \bibinfo {pages} {1480} (\bibinfo {year} {2004})}\BibitemShut {NoStop}%
\bibitem [{\citenamefont {Heinen}\ \emph {et~al.}(2011)\citenamefont {Heinen},
  \citenamefont {Holmqvist}, \citenamefont {Banchio},\ and\ \citenamefont
  {N\"{a}gele}}]{heinen:yukawa}%
  \BibitemOpen
  \bibfield  {author} {\bibinfo {author} {\bibfnamefont {M.}~\bibnamefont
  {Heinen}}, \bibinfo {author} {\bibfnamefont {P.}~\bibnamefont {Holmqvist}},
  \bibinfo {author} {\bibfnamefont {A.~J.}\ \bibnamefont {Banchio}}, \ and\
  \bibinfo {author} {\bibfnamefont {G.}~\bibnamefont {N\"{a}gele}},\ }\href
  {\doibase 10.1063/1.3524309} {\bibfield  {journal} {\bibinfo  {journal} {J.
  Chem. Phys.}\ }\textbf {\bibinfo {volume} {134}},\ \bibinfo {eid} {044532}
  (\bibinfo {year} {2011})}\BibitemShut {NoStop}%
\bibitem [{\citenamefont {Asakura}\ and\ \citenamefont
  {Oosawa}(1958)}]{asakura:asakuraoosawa}%
  \BibitemOpen
  \bibfield  {author} {\bibinfo {author} {\bibfnamefont {S.}~\bibnamefont
  {Asakura}}\ and\ \bibinfo {author} {\bibfnamefont {F.}~\bibnamefont
  {Oosawa}},\ }\href {\doibase 10.1002/pol.1958.1203312618} {\bibfield
  {journal} {\bibinfo  {journal} {J. Polym. Sci.}\ }\textbf {\bibinfo {volume}
  {33}},\ \bibinfo {pages} {183} (\bibinfo {year} {1958})}\BibitemShut
  {NoStop}%
\bibitem [{\citenamefont {Roth}, \citenamefont {Evans},\ and\ \citenamefont
  {Dietrich}(2000)}]{roth:asakuraoosawa}%
  \BibitemOpen
  \bibfield  {author} {\bibinfo {author} {\bibfnamefont {R.}~\bibnamefont
  {Roth}}, \bibinfo {author} {\bibfnamefont {R.}~\bibnamefont {Evans}}, \ and\
  \bibinfo {author} {\bibfnamefont {S.}~\bibnamefont {Dietrich}},\ }\href
  {\doibase 10.1103/PhysRevE.62.5360} {\bibfield  {journal} {\bibinfo
  {journal} {Phys. Rev. E}\ }\textbf {\bibinfo {volume} {62}},\ \bibinfo
  {pages} {5360} (\bibinfo {year} {2000})}\BibitemShut {NoStop}%
\bibitem [{\citenamefont {Yan}\ \emph {et~al.}(2006)\citenamefont {Yan},
  \citenamefont {Buldyrev}, \citenamefont {Giovambattista}, \citenamefont
  {Debenedetti},\ and\ \citenamefont {Stanley}}]{yan:ramp}%
  \BibitemOpen
  \bibfield  {author} {\bibinfo {author} {\bibfnamefont {Z.}~\bibnamefont
  {Yan}}, \bibinfo {author} {\bibfnamefont {S.~V.}\ \bibnamefont {Buldyrev}},
  \bibinfo {author} {\bibfnamefont {N.}~\bibnamefont {Giovambattista}},
  \bibinfo {author} {\bibfnamefont {P.~G.}\ \bibnamefont {Debenedetti}}, \ and\
  \bibinfo {author} {\bibfnamefont {H.~E.}\ \bibnamefont {Stanley}},\ }\href
  {\doibase 10.1103/PhysRevE.73.051204} {\bibfield  {journal} {\bibinfo
  {journal} {Phys. Rev. E}\ }\textbf {\bibinfo {volume} {73}},\ \bibinfo
  {pages} {051204} (\bibinfo {year} {2006})}\BibitemShut {NoStop}%
\bibitem [{\citenamefont {Jagla}(1999)}]{jagla:ramp}%
  \BibitemOpen
  \bibfield  {author} {\bibinfo {author} {\bibfnamefont {E.~A.}\ \bibnamefont
  {Jagla}},\ }\href {\doibase 10.1063/1.480241} {\bibfield  {journal} {\bibinfo
   {journal} {J. Chem. Phys.}\ }\textbf {\bibinfo {volume} {111}},\ \bibinfo
  {pages} {8980} (\bibinfo {year} {1999})}\BibitemShut {NoStop}%
\bibitem [{\citenamefont {Errington}, \citenamefont {Truskett},\ and\
  \citenamefont {Mittal}(2006)}]{errington:ramp}%
  \BibitemOpen
  \bibfield  {author} {\bibinfo {author} {\bibfnamefont {J.~R.}\ \bibnamefont
  {Errington}}, \bibinfo {author} {\bibfnamefont {T.~M.}\ \bibnamefont
  {Truskett}}, \ and\ \bibinfo {author} {\bibfnamefont {J.}~\bibnamefont
  {Mittal}},\ }\href {\doibase 10.1063/1.2409932} {\bibfield  {journal}
  {\bibinfo  {journal} {J. Chem. Phys.}\ }\textbf {\bibinfo {volume} {125}},\
  \bibinfo {eid} {244502} (\bibinfo {year} {2006})}\BibitemShut {NoStop}%
\bibitem [{\citenamefont {Hlushak}, \citenamefont {Hlushak},\ and\
  \citenamefont {Trokhymchuk}(2013)}]{hlushak:sexpfmsasquareshoulder}%
  \BibitemOpen
  \bibfield  {author} {\bibinfo {author} {\bibfnamefont {S.~P.}\ \bibnamefont
  {Hlushak}}, \bibinfo {author} {\bibfnamefont {P.~A.}\ \bibnamefont
  {Hlushak}}, \ and\ \bibinfo {author} {\bibfnamefont {A.}~\bibnamefont
  {Trokhymchuk}},\ }\href {\doibase 10.1063/1.4801659} {\bibfield  {journal}
  {\bibinfo  {journal} {J. Chem. Phys.}\ }\textbf {\bibinfo {volume} {138}},\
  \bibinfo {eid} {164107} (\bibinfo {year} {2013})}\BibitemShut {NoStop}%
\bibitem [{\citenamefont {Tang}\ and\ \citenamefont {Lu}(1997)}]{tang:fmsa2}%
  \BibitemOpen
  \bibfield  {author} {\bibinfo {author} {\bibfnamefont {Y.}~\bibnamefont
  {Tang}}\ and\ \bibinfo {author} {\bibfnamefont {B.~C.-Y.}\ \bibnamefont
  {Lu}},\ }\href {\doibase 10.1080/002689797172697} {\bibfield  {journal}
  {\bibinfo  {journal} {Mol. Phys.}\ }\textbf {\bibinfo {volume} {90}},\
  \bibinfo {pages} {215} (\bibinfo {year} {1997})}\BibitemShut {NoStop}%
\bibitem [{\citenamefont {Tang}(2003)}]{tang:fmsa}%
  \BibitemOpen
  \bibfield  {author} {\bibinfo {author} {\bibfnamefont {Y.}~\bibnamefont
  {Tang}},\ }\href {\doibase 10.1063/1.1541615} {\bibfield  {journal} {\bibinfo
   {journal} {J. Chem. Phys.}\ }\textbf {\bibinfo {volume} {118}},\ \bibinfo
  {pages} {4140} (\bibinfo {year} {2003})}\BibitemShut {NoStop}%
\end{thebibliography}%

\end{document}